# Fully on-chip microwave photonics system

Yuansheng Tao[1,†], Fenghe Yang[4,†], Zihan Tao[1], Lin Chang[1,2], Haowen Shu[1], Ming Jin[1], Yan Zhou[4], Zhangfeng Ge[4], and Xingjun Wang[1,2,3,4,*]

[1] State Key Laboratory of Advanced Optical Communications System and Networks, School of Electronics, Peking University, Beijing 100871, China.

[2] Frontiers Science Center for Nano-optoelectronics, Peking University, Beijing 100871, China.

[3] Peng Cheng Laboratory, Shenzhen 518055, China.

[4] Peking University Yangtze Delta Institute of Optoelectronics, Nantong 226010, China.

*Corresponding author: xjwang@pku.edu.cn

† These authors contributed equally to this manuscript

**Microwave photonics (MWP), harnessing the tremendous bandwidth of light to generate, process and measure wideband microwave signals, are poised to spark a new revolution for the information and communication fields. Within the past decade, new opportunity for MWP has emerged driven by the advances of integrated photonics. However, despite significant progress made in terms of integration level, a fully on-chip MWP functional system comprising all the necessary photonic and electronic components, is yet to be demonstrated. Here, we break the status quo and provide a complete on-chip solution for MWP system, by exploiting hybrid integration of indium phosphide, silicon photonics and complementary metal-oxide-semiconductor (CMOS) electronics platforms. Applying this hybrid integration methodology, a fully chip-based MWP microwave instantaneous frequency measurement (IFM) system is experimentally demonstrated. The unprecedented integration level brings great promotion to the compactness, reliability, and performances of the overall MWP IFM system, including a wide frequency measurement range (2-34 GHz), ultralow estimation errors (10.85 MHz) and a fast response speed (∼0.3 ns). Furthermore, we deploy the chip-scale MWP IFM system into realistic application tasks, where diverse microwave signals with rapid-varying frequencies at X-band (8-12 GHz) are accurately identified in real-time. This demonstration marks a milestone for the development of integrated MWP, by providing the technology basis for the miniaturization and massive implementations of various MWP functional systems.**

## Introduction

With the impending electronic bandwidth bottleneck in our information society for radio-frequency (RF) networks and Internet of Things [1], the use of photonics to generate, process and measure wideband microwave signals has been explored extensively, which nowadays is well known as microwave photonics (MWP) [2,3]. Its large bandwidth and low-loss features enable the realization of key functionalities in microwave systems that are not offered by current RF technology. Recent years, in particular, new opportunity for MWP has emerged with the advances of integrated photonic circuit (PIC) [4,5], which enables a dramatic reduction in the system size, weight, and power-consumption (SWaP). Milestones have been achieved in a wide range of MWP functionalities, including filters [6–8], arbitrary waveform generators [9,10], microwave frequency measurement [11–13], phase shifters [14], tunable true-time delay lines [15], beamformers [16], and generic programmable processors [17–19], exhibiting a various level of system-integration completeness.

However, despite significant progress, a complete on-chip solution to fully incorporate all the required photonic components and supporting electronic hardware of MWP system is still missing: currently, most of MWP systems have only implemented partial photonic components as chip-integrated format [8–19], while the rest of the system are still constituted by bulk devices or equipment, thereby will induce issues related to large footprint, poor robustness, and high power-consumption. While indium phosphide (InP) platform allows monolithic integration of all photonic elements [6], it suffers from relatively large passive waveguide loss and elevated amplifier noise [5]. Silicon photonics (SiPh) platform is attractive for its scalable and low-cost implementation of diverse building blocks of MWP systems [20], but the integration of light source requires significant extra efforts. Other integrated photonic materials, such as silicon nitride [21], chalcogenide [22], and lithium niobate [23] also encounter similar obstacles in high-level integration. Therefore, finding a versatile path that combines strength from diverse material platforms is a prior task for the goal of a fully on-chip MWP system.

Aside from the photonic integration, another critical issue is the convergence with electronics integration that has been missing from most of the integrated MWP works. Generally, to offer powerful RF functionalities, many MWP applications should leverage high-performance electronic devices in addition to the photonic parts [4,24]. For instance, in an optoelectronic oscillator (OEO) [25,26], high-gain modulator drivers should be adopted to compensate the optical insertion loss for inspiring the microwave oscillation. In a microwave frequency measurement system [27], lock-in amplifiers are capable to enhance the readout accuracy of weak optical signals. However, the electronic hardware existed in current integrated MWP systems is still being implemented by large-size discrete modules, deeply limiting the scalability, efficiency, and noise performances. Hence, exploiting the integration with modern complementary metal-oxide-semiconductor (CMOS) electronics [28], is also crucial to the miniaturization of MWP systems.

Here, we demonstrated the first complete on-chip solution for MWP system leveraging hybrid integration of different photonic platforms (InP and SiPh) and CMOS electronic circuits. Applying this integration strategy, a fully chip-integrated MWP instantaneous frequency measurement (IFM) system is achieved. The photonic integration is implemented based on directly coupling InP distributed feedback (DFB) laser with SiPh PIC that is compactly united with CMOS transimpedance amplifier (TIA) dies using wire-bonding. Thanks to the unprecedented integration level, the MWP IFM system makes a huge leap compared to the prior studies with similar functionality [11–13,29–34], in terms of small footprint (tens of mm$^2$), wide frequency measurement range (2-34 GHz), ultralow estimation error (10.85 MHz) and fast response speed (~0.3 ns). Also, the key capacity of real-time frequency identification on rapid-varying microwave signals, is experimentally performed at X-band (8-12 GHz) with single-shot detection and high accuracy. Such complete on-chip solution in this work can find immediate practices in numerous MWP operational systems across the microwave signal generation [35], transmission [36], and processing [37] domains.

## Results

### MWP system architecture and principle

To prove the feasibility of the proposed complete chip-integrated solution, we experimentally implement a fully on-chip MWP functional system for microwave frequency measurement application, as a concrete paradigm. The architecture of the MWP IFM system is illustrated in Fig. 1a, consisting of an InP DFB laser, a monolithic SiPh PIC and electronic CMOS TIAs. The MWP IFM system performs microwave frequency identification based on frequency to optical power mapping approach [38]. Firstly, a continuous-wave (c.w.) optical carrier is generated by the InP DFB laser, and then injected into the SiPh PIC via butt-coupling. The microwave signal of unknown instantaneous frequency ($f_{RF}$) is modulated onto the optical carrier by a high-speed Mach-Zehnder (MZ) modulator. The MZ modulator is driven as push-pull scheme and biased at the minimum transmission point to generate a dual-sideband suppressed-carrier (DSB-SC) optical modulated signal. The main reason for selecting DSB-SC modulation is due to its simple generation (only need single bias control) and low power dissipation, thereby is suitable for system-level integration. Noted that the bias-drift of silicon MZ modulator is relatively small [39] compared to those of bulky lithium-niobate modulators used in previous studies, therefore, the produced DSB-SC signal is highly stable during operation. A micro-ring resonator (MRR) with high-Q factor (narrow stopband) is followed to further remove the residual optical carrier. Subsequently, the DSB-SC signal is routed into an asymmetrical MZ interferometer (AMZI) filter, which serves as the key linear-optics frequency discriminator in our MWP IFM system. Once the optical carrier wavelength is finely aligned to the central transmission spectra of the bar/cross ports, thereby, decided by the inherently complimentary responses of AMZI, the DSB-SC signals will undergo $f_{RF}$-dependent attenuation with opposite slopes at the bar/cross out-ports (as shown in label iv in Fig. 1a). The two channel DSB-SC optical signals are detected in a pair of germanium (Ge)/Si PDs. The resulting weak photocurrents are amplified and converted into voltage-format waveform by dual TIAs, and finally collected by a real-time oscilloscope for post-processing. Assuming the two output electrical powers are severally denoted by $P_{bar}(f_{RF})$ and $P_{cross}(f_{RF})$, in this case, the power ratio $P_{bar}/P_{cross}$ shall be an unambiguous function of input microwave frequency ($f_{RF}$) within the frequency range of half free spectral range (FSR) of the AMZI discriminator, which is generally known as the amplitude comparison function (ACF) [29], given as below:

$$ACF(f_{RF}) = \frac{P_{bar}(f_{RF})}{P_{cross}(f_{RF})} \quad (1)$$

Based on this frequency-to-power mapping ACF, the unknown frequency of input microwave signal can be estimated. More detailed theoretical analysis is provided in Supplementary Note 1.

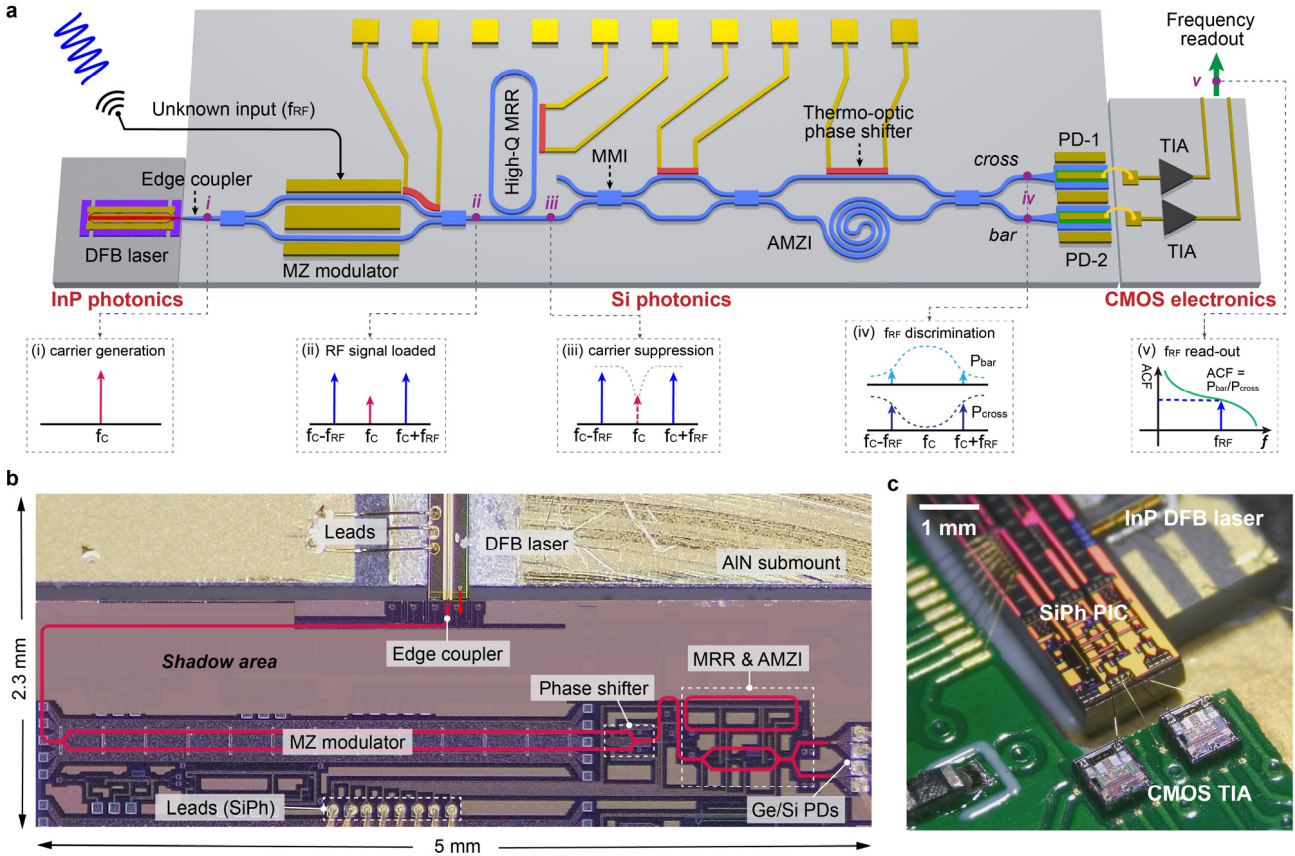

**Fig. 1 Photonic-electronic fully on-chip MWP IFM system. a,** Illustration and operation principles of the fully chip-integrated MWP IFM system. **b,** Top-view micrograph of the hybrid-integrated InP DFB laser and SiPh PIC. Noted that the highlighted beam-path of MRR&AMZI is roughly showed, due to their relatively small size. Details can be seen in Fig. 2e and f. **c,** Photograph of the implemented fully on-chip MWP IFM system. The SiPh PIC is directly wire-bonded to the electronic CMOS TIA chips, co-packaged on a PCB.

**Design and device characterization**

Fig. 1b and c show the photographs of the MWP IFM system, that is fully implemented as chip-integrated formats and co-packaged onto a custom-designed print circuit board (PCB) for electrical connections. Fig. 2a-f display the essential photonic and electronic building units of this MWP system and their key performance metrics. For c.w. light generation, a commercially available InP DFB laser diode chip is employed, with a tunable emission wavelength centered at 1552.3 nm (see Fig. 2a). The maximum output power is measured as ~100 mW under a pumping current of 300 mA (Supplementary Note 2). The monolithic SiPh PIC is designed in-house and fabricated by CompoundTek Foundry Services using its standard 90-nm silicon-on-insulator (SOI) lithography process. The InP and SiPh PICs are butt connected based on a Si inverse-taped edge coupler. The facet-to-facet coupling loss is measured as about 6.05±0.35 dB (Supplementary Note 3). The Si MZ modulator (3-mm long) is implemented based on travelling-wave PN depletion scheme, showing a modulation bandwidth of >23 GHz (see Fig. 2c). Fig. 2d shows the vertical p-doped-intrinsic-n-doped (PIN) Ge/Si PD, with ultralow dark photocurrents of ~9 nA, thus supporting a large detection dynamic range. The MRR is designed as multimode waveguide structures to improve the Q-factor (~$1.4×10^5$, shown in Fig. 2e), and the coupling coefficients can be adjusted in virtue of the interferometric scheme (details are provided in Supplementary Note 4) to access critical coupling state. In this sense, the optimized MRR enables nearly ideal carrier suppression of the DSB-SC modulated signals, that is high rejection ratio and slight impact on close-to-carrier sidebands. For the AMZI discriminator, an extra interferometer function as variable beam splitter, is placed at the front (see in Fig. 1a) to form an adaptive double-MZI (DMZI) structure [40]. This particular design is to compensate the imperfections of fabricated 1×2/2×2 multi-mode interference (MMI) splitters and the unbalanced propagation loss of two AMZI arms, further to avoid the degradation of extinction ratio (corresponding to the measurement accuracy). The transmission spectrum of AMZI is characterized (in Fig. 2f), showing an FSR of 0.74 nm (92.5 GHz). TiN thermo-optic (TO) phase shifters are used to tune and maintain the desired operating bias or status for the above-mentioned SiPh devices. The SiPh PIC is directly wire bonded to dual unpackaged commercial TIA dies fabricated in a standard CMOS process (see in Supplementary Note 5). The maximal gain is measured as 34.8 dB at 10 MHz (in Fig. 2b). More details about the performance characterization can be found in Methods.

Thanks to the fully chip-scale integration of all required components as noted above, compared with the state-of-the-art benchtop or partially integrated MWP IFM systems, our MWP IFM system only operates at a fraction of the power consumption (884.2 mW, see in Supplementary Note 6) and requires a drastically reduced footprint (Fig. 1c, tens of $mm^2$). These features make it very suitable for mass production and widespread use in some emerging applications (e.g., airborne systems) in near future.

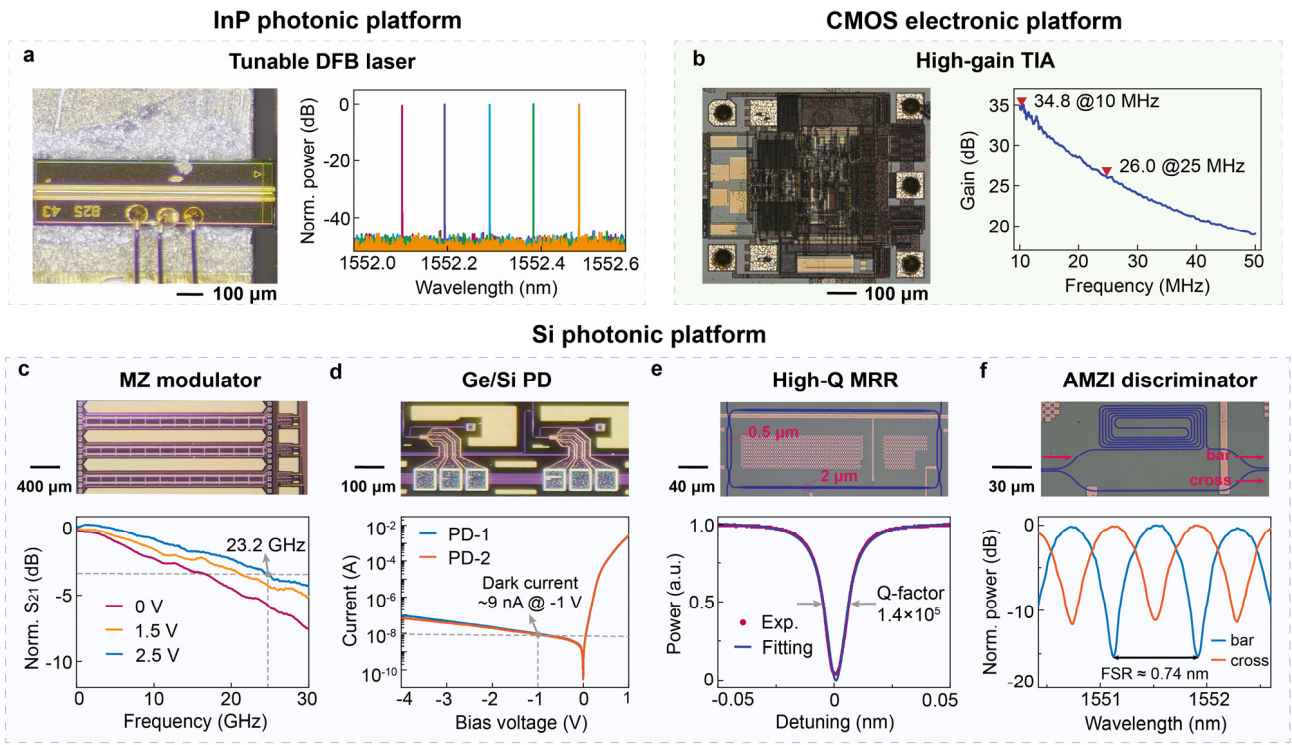

**Fig. 2 Key photonic/electronic devices and fundamental characteristics. a,** InP DFB laser with a tunable emission wavelength around 1552 nm. **b,** CMOS-based TIA for weak current amplification

after Ge/Si PD. The maximal gain at 10 MHz is 34.8 dB. **c,** High-speed silicon MZ modulator. The 3-dB operation bandwidth is > 23 GHz. **d,** Vertical PIN Ge/Si PDs, both have a low dark current of ~9 nA at -1 V bias. **e,** High-Q MRR used to perform carrier suppression, with a loaded Q-factor of $1.4 \times 10^5$. **f,** AMZI discriminator with an FSR of 0.74 nm, function as the optical discriminator in the MWP IFM system.

**Static microwave frequency identification**
The static microwave frequency identification capability of the MWP IFM system is firstly evaluated, in which the input is c.w. microwave signal with constant frequencies. Owing to the full photonic integration, the experimental setup (shown in Fig. 3a) is quite simple, while the only employed discrete equipment is just as tools for signal generation and characterization, and is not a part of the MWP IFM system. Noted that the CMOS TIAs are not used in this static measurement, since the resulting photocurrents are dc signals.

To initiate the frequency identification operation, we start with calibrating the bias of MZ modulator and the status of high-Q MRR, to realize the generation of DSB-SC modulated signal as shown in Fig. 3b. With the incorporation of the high-Q MRR filtering, the obtained optical carrier suppression ratio is higher than 30 dB, and can be maintained across broad frequency range (Supplementary Note 7). Then, the integrated MWP IFM system is characterized by sweeping the input microwave frequencies across 2-35 GHz and recording the generated photocurrents at the pair of on-chip Ge/Si PDs. The measured results are normalized and shown in the upper panel of Fig. 3c. As expected, the electrical responses of two Ge/Si PDs show opposite trends with frequency, in conformity with the AMZI complementary transmission spectra multiplied by frequency-dependent roll-off of the MZ modulator. It can be noticed that slight ripples exist in the PD frequency responses, which mainly derive from the power fluctuations of the free-running DFB laser. Nonetheless, by making subtraction of these two PD response curves based on Eq. (1), the power ripples can be effectively eliminated for a smooth ACF profile (lower panel of Fig. 3c), because the variations of input power will create exactly same impacts on PD-1 (bar port) and PD-2 (cross port). Following, a polynomial method is adopted to fit the measured data for the realization of continuous ACF curve. The obtained ACF curve, in turn, is used to estimate unknown input microwave frequencies in the whole measurement range, as shown in Fig. 3d. The figure indicates that the MWP IFM system is capable of identifying frequencies with very high precision up to 34 GHz. The estimation errors increased at high frequencies (>34 GHz), seen as inset figure of Fig. 3d, are mainly resulted by the limited EO bandwidth of silicon modulator (~23 GHz in our system). Meanwhile, the near-dc frequency (<2 GHz) inputs are influenced by the MRR filtering profile and also cannot be accurately estimated. Therefore, the valid high-accuracy frequency measurement range is around 2-34 GHz. Within this measurement range, the frequency estimation errors have an ultralow root mean square (RMS) value of only 10.85 MHz (see in Fig. 3e). Moreover, the long-term stability is also evaluated by monitoring the average estimation errors over 0.5-hour experiment, as shown in Fig. 3f, and the results show acceptable fluctuations (<12.5%) on identification accuracy. More details about the experiment are given in Methods.

An important figure of merit (FOM) for microwave IFM systems is the estimation error as a percentage of the measurement range [11], which is calculated as 0.033% in this work. To the best of our knowledge, the attained FOM is a record-low value among linear-optics MWP IFM systems. This remarkable achievement is attributed to the complete integration of the photonic link, which accordingly leads to great improvement on the robustness for environment perturbations. While in this regards, former partially integrated MWP IFM systems usually require tens of centimeters or longer fibers to connect the on-chip optical discriminator and other off-chip devices, which will be inevitably influenced by operating variations (e.g., the polarization, vibration, temperature) and thereby cause estimation errors. In addition, the full photonic integration avoids the extra light amplification to compensate coupling loss, which itself will add amplified spontaneous emission noise into the system link.

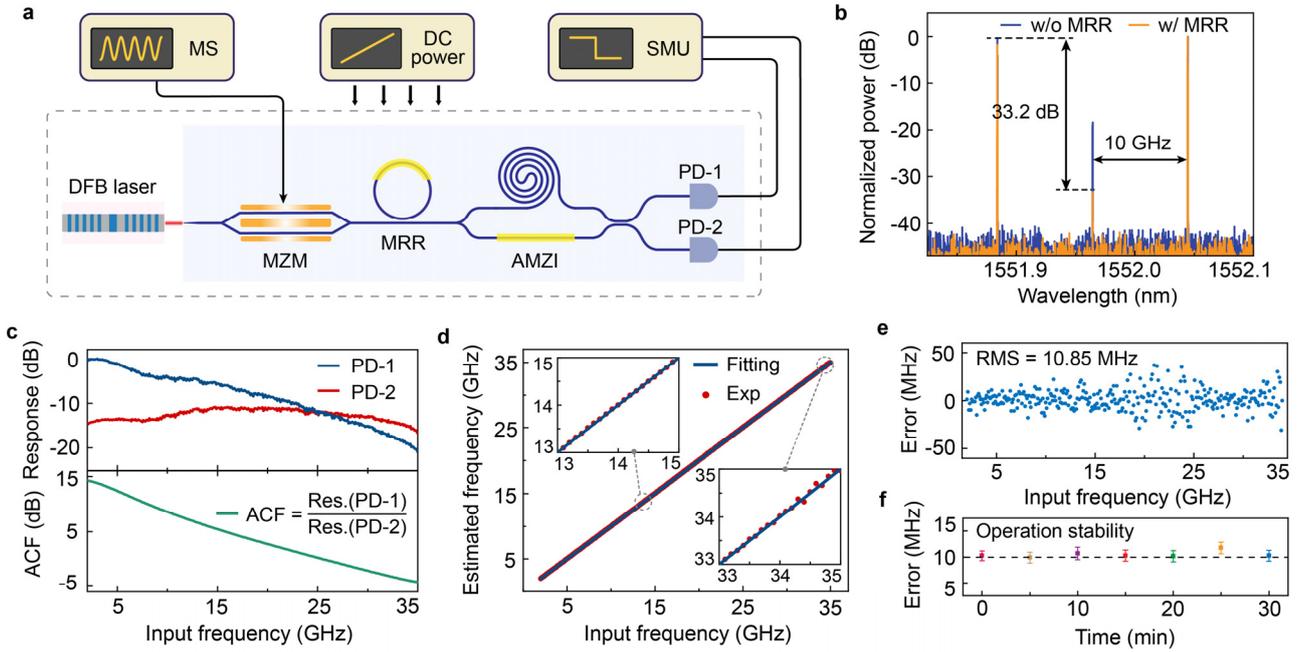

**Fig. 3 Static microwave frequency identification using the integrated MWP IFM system. a,** Experimental setup to perform the microwave frequency identification as static manner. MS, microwave source; SMU, source/measure unit; MZM, MZ modulator. **b,** The generated optical DSB-SC modulation signal without and with the aid of high-Q MRR suppression, at a modulation frequency of 10 GHz. **c,** Upward, electrical responses of the pair of Ge/Si PDs; below, the corresponding ACF curve. **d,** Frequency estimation measurement over 2-35 GHz. Inset: Enlarged estimation results at the range of 13-15 GHz and 33-35 GHz. **e,** Frequency estimation errors over 2-34 GHz, showing an RMS of 10.85 MHz. **f,** Characterization of long-term stability. Measured mean absolute estimation error over 30 minutes.

## Real-time microwave frequency identification

Besides static microwave frequency measurement, a broad range of practical IFM application scenarios involving radar warning receiver [29], biomedical instrument [41] and cognitive radio [42], commonly demand the identification of dynamic microwave signals with time-varying instantaneous frequencies. To validate that our fully on-chip MWP IFM system also holds the real-time frequency identification capacity, we apply it to recognize several typical time-varying microwave signals including hopping-frequency (HF), linear chirped-frequency (CF) and quadratic CF signals. The experimental setup is shown in Fig. 4a, and details are provided in Methods.

Fig. 4b lists the temporal frequency sequences of input microwave signals with µs-level variation rate, generated by an arbitrary waveform generator (AWG, 50 GSa/s). For the HF signal, the frequency sequences hop in the following order: 8.2, 10.6, 9.1, 11.7, 7.5 and 9.8 GHz, each with the duration of 1 µs. For the linear/quadratic CF signals, the chirping frequency range is 7.5-11.5 GHz, with a total chirping duration of 10 µs. When processed through the fully on-chip MWP IFM system, these input time-varying microwave oscillations (~GHz) are broadcasted onto optical carrier and discriminated based on frequency-to-power mapping in real time, and ultimately the targeted frequency variations with relatively low-speed (~MHz) will be extracted back to electrical domain. The bandwidth of Ge/Si PDs (~31 GHz, Supplementary Note 8) and CMOS TIAs (>10 MHz, Fig. 2b) are adequate to recognize the frequency changing rates of these incoming microwave signals. To recover the input frequency sequences, we only need a low-bandwidth oscilloscope to record the temporal output responses of the MWP IFM system, and perform simple off-line processing based on the pre-measured ACF curve shown in Fig. 4e. Fig. 4c displays the real-time readout frequency variation profiles of the HF, linear CF, and quadratic CF microwave signals, respectively. Compared with the original inputs (Fig. 4b), it can be seen that these three distinctly different time-varying microwave signals are all identified with high accuracy. The histograms of estimation errors are given in Fig. 4d. The corresponding RMS is calculated as 60.5 MHz, 59.8 MHz, and 55.0 MHz, respectively. These achieved RMS values are a little larger than that of static frequency identification, which mainly arises from the additional noise of CMOS TIAs, but still better than the measurement accuracy (in static manner) of most previous works. The input frequencies and their variation rates in this experiment are only decided by the off-the-shelf AWG (sampling-rate limited) and TIA dies (bandwidth limited), not the entire capacities of the MWP IFM system. To explore the fastest respond speed for instantaneous frequency-bursts, a HF signal with hopping frequency sequence of 7-12 GHz is chosen as test input (Fig. 4f), and in this case, a discrete low-noise electrical amplifier with higher bandwidth is employed instead of the TIA chips. The corresponding temporal amplitude-stepped waveform is measured and shown in Fig. 4g, indicating an ultra-short transition time of ~0.3 ns.

It worth emphasized here that the real-time high-accuracy microwave frequency identifications demonstrated in this work are realized by single-shot detection without averaging processing (used in previous work [13]). This is due to the uniting of the SiPh PIC with micro-electronics TIAs, that enables the signal to noise ratio (SNR) of the overall MWP system to be greatly improved. Moreover, our work also avoids the cumbersome frequency-scanning operation [12], thereby reduces the implementation cost and complexity. Therefore, based on its operating simplicity and complete system integration, the fully on-chip MWP IFM system bridges the gap between laboratory test and real-world deployments, and the further efforts are supposed to be focus on the optimizations of element devices and packaging, which will be discussed in later section.

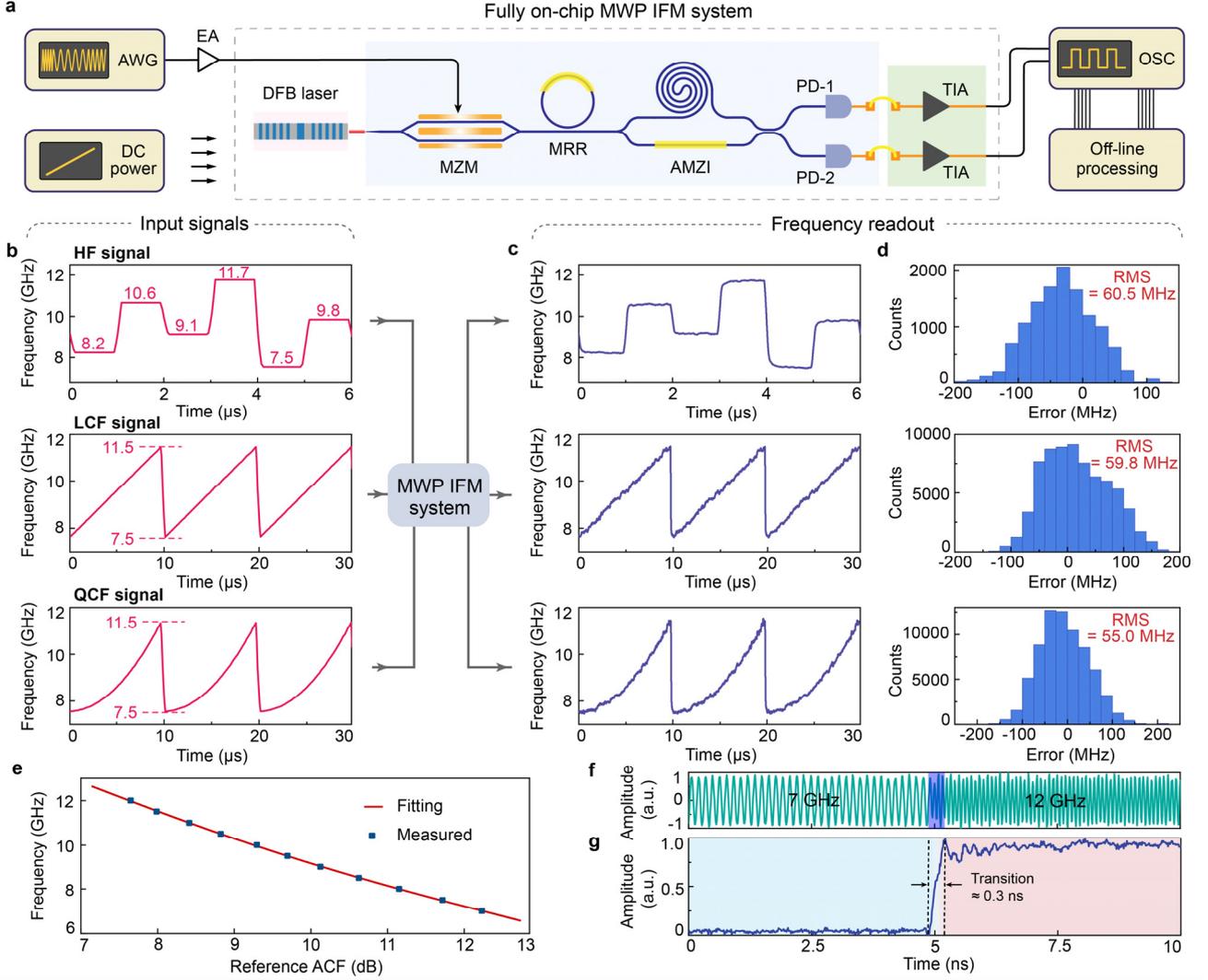

**Fig. 4 Real-time microwave frequency identification using the fully on-chip MWP IFM system.**
**a,** Experimental setup to perform the real-time microwave frequency identification. AWG, arbitrary waveform generator; OSC, oscilloscope. **b,** Input dynamically time-varying microwave signals, including HF signal, linear-CF (LCF) signal, and quadratic-CF (QCF) signal. **c,** Real-time frequency readout processed through the integrated MWP IFM system. **d,** Histogram of the estimation errors. **e,** The reference ACF employed in these real-time microwave frequency identification tests. **f,** The time-domain waveform of a HF signal with 7 GHz-12 GHz hopping sequences, used to explore the shortest transition time. **g,** The temporal response of the integrated MWP IFM system, showing a transition time of ~0.3 ns.

## Discussion

A comparison among the representative integrated MWP IFM systems is shown in Table I. We see that the overall frequency identification performance presented in this work is a great advance compared to the pioneering studies, due to the combination of wide measurement range, high accuracy, and fast response speed. This advancement is attributed to the record-high system integration containing all the required photonic and electronic devices, thereby enabling the compactness, measurement latency, reliability, and noise performance of the MWP IFM system to be significantly improved. Noted that though stimulated Brillouin scattering (SBS)-based IFM systems may show lower estimation error [12], they use a large pumping power (>100 mW) and complex frequency-scanning technique (defies real-time frequency identification capacity), which will hinder their practical deployments in many real-world scenarios.

**Table I.** Comparison of state-of-the-art integrated MWP IFM systems

| Ref | System integration | | | | Measurement Range (GHz) | Error (MHz) | FOM[b] | Response time (ns) |
|---|---|---|---|---|---|---|---|---|
| | Passive | Active[a] | Light source | Electronics | | | | |
| [30] | Y | N | N | N | 0.5-4 | 93.6 | 2.7 | N/A |
| [31] | Y | N | N | N | 3-19 | 500 | 3.1 | N/A |
| [32] | Y | N | N | N | 1-30 | 237 | 0.8 | ~1×10$^7$ |
| [33] | Y | N | N | N | 0-26.62 | 250 | 0.9 | N/A |
| [34] | Y | N | N | N | 5-20 | 47.2 | 0.3 | ~2×10$^8$ |
| [11] | Y | N | N | N | 0-40 | 319 | 0.8 | N/A |
| [12] | Y | N | N | N | 9-38 | 1 | 0.003 | ~6×10$^5$ |
| [13] | Y | N | N | N | 0.01-32 | 755 | 2.3 | 1 |
| This work | Y | Y | Y | Y | 2-34 | 10.85 | 0.03 | 0.3 |

[a] refers to high-speed modulator and PD

[b] represents the estimation error as a percentage (%) of the measurement range.

In current implementation, the hybrid integration of InP platform and Si platform is realized based on butt-coupling, which can be replaced by photonic wire bonding technique [43] for more efficient optical connection. Heterogeneous integration [44] can further improve the compactness and scalability, but it still needs significant efforts in manufacturing development. Moving forward, the fundamental photonic devices also have the potential for upscaling, toward a more powerful MWP IFM system. For example, the valid frequency measurement range can be enlarged by optimizing the EO bandwidth of silicon modulator (potentially >50 GHz [45]). To promote the identification accuracy, the key is to access a broadband and ultra-sharp ACF profile, which is possible to be satisfied by utilizing Fano resonance structures [46]. On the electronic side, by virtue of the excellent CMOS compatibility of silicon, the low-noise TIAs are capable of being seamlessly integrated with photonics on a monolithic platform [47], providing solutions to drastically reduce the production cost.

Beyond the specific microwave frequency measurement demonstration provided here, our photonic-electronic complete on-chip solution also holds a profound significance to benefit much broader range of MWP applications, due to the universality of this work. For example, in OEO systems [35], the fully chip-scale integration will dramatically improve the power efficiency, stability of oscillation frequency and the key phase noise performance metrics. In the microwave signal processing such as the photonic analog-to-digital converter (ADC) [48] and channelized receiver [49], they usually demand an array of digital electronic circuits for post-processing. Implementing the hybrid photonic-electronic integration will support the reduction in footprint and promotion in the signal-to-noise ratio of the overall system. Therefore, we anticipate our demonstration could serve as a general strategy for the further development of integrated MWP, and paves the way for the massive applications of this technology in next generation information and communication fields.

## Methods

**Characterizations of the key photonic/electronic devices.** To characterize the performances of diverse photonic devices in the MWP system, monitoring ports are preset in several critical positions of the entire on-chip optical link using 1:9 directional couplers. Firstly, the emission spectrum of InP DFB laser is collected and characterized by an ultrahigh-resolution (10 MHz) optical spectrum analyzer (Aragon Photonics, BOSA300 C+L), exhibiting a very narrow spectral linewidth and >40 dB side-mode suppression ratio. The electro-optic (EO) bandwidth of MZ modulator is measured by a vector network analyzer (VNA, Keysight N5247A), with the typical results of >23 GHz (at 2.5 V reverse bias voltage). For the waveguide-coupled Ge/Si PDs, the dark photocurrent is the most concern performance metrics in our IFM application, which is measured by a high-precision source meter (Keysight B2902A) under different bias voltages (using the sweep function mode). For the accurate evaluation of passive MRR and AMZI, a tunable laser (Keysight 81960A) is used to sweep the wavelength, meanwhile the transmission of optical signal is converted to electrical signal by a photodetector and then received by an oscilloscope (Agilent DSO81204B) synchronized with the laser. The CMOS TIA for signal amplification is characterized in terms of its gain value at different input frequency, that is tested by a VNA (Keysight N5247A). Due to the restriction of the measurement equipment, we could only obtain the gain values at >10 MHz range.

**Static microwave frequency measurement experiment.** The InP-based DFB laser is driven at 300 mA pumping current to generate ~100 mW c.w. optical carrier, employing a commercial laser diode controller (Thorlabs, ITC 4001). The c.w. optical carrier is launched into the SiPh chip as butt-coupling manner, with a coupling loss of ~6 dB. Both the InP and SiPh chips are mounted on thermoelectric cooler to stabilize the temperature during operation. An analog signal generator (Keysight E8257D) is employed to produce 16 dBm microwave signals of different frequencies, as input for the integrated IFM system. The microwave signals to be measured are firstly sent into a 180° RF hybrid (Marki); and then the out-of-phase dual outputs are fed to the inputs of silicon MZ modulator using coaxial cables of equal length. Considering the parasitic inductance of Au bonding wires, to support high-bandwidth IFM operation, in this test, the RF connection to the on-chip MZ modulator is implemented by GSGSG high-speed RF probes (Cascade, ACP40). The DC calibrations of TO phase shifters are offered by multi-channel DC power supplies (Keysight E36312A). The output photocurrents of the on-chip Ge/Si PDs are simultaneously collected by a two-channel high-precision source meter unit (Keysight B2902A). The analog signal generator and the source meter are programmatically controlled in MATLAB environment on a desktop via general purpose interface bus (GPIB) interfaces, which enables a fast and accurate measurement.

**Real-time microwave frequency measurement experiment.** The basic experimental setup is same with that of static IFM demonstration, while the major differences lie in the signal transmitting and receiving parts. In the transmitting part, a RF arbitrary waveform generator (AWG,

Tektronix AWG70001) with a sampling rate of 50 GSa/s, is employed to generate the rapidly time-varying HF, linear CF, and quadratic CF microwave signals. The produced signals of the AWG are firstly amplified by a linear electrical amplifier (Mini-circuits), and then are fed into the RF input of the fully on-chip MWP IFM system, or collected by a 39 GHz real-time oscilloscope (Teledyne LeCroy, MCM-Zi-A) for recording original RF temporal waveforms as references (as displayed in Fig. 4b after Fourier transformation). In the receiving part, the generated temporal waveforms that contain the desired time-varying frequency sequences, are collected by a low-speed real-time oscilloscope (RIGOL DS7014, 10 GSa/s). A frequency step sequence 7.5-0.5-12.5 GHz is chosen as input, to obtain the reference ACF curve of the MWP IFM system (Fig. 4e). The data are analyzed offline using MATLAB on a desktop. In addition, to explore the shortest transition time for instantaneous frequency-burst, a HF signal (hopping at 7, 12 GHz) is chosen as test input (Fig. 4f). Since our off-the-shelf TIA dies are bandwidth-limited, in this case, we amplify the photocurrents using a low-noise electrical amplifier (LNA, Mini-circuits) with higher bandwidth and collect time-domain response by a 39 GHz oscilloscope (Teledyne LeCroy, MCM-Zi-A).

**Acknowledgment**
This work was supported by National Key R&D Program of China (Grant No. 2021YFB2800400 and 2021YFB2801200), Beijing Natural Science Foundation (Z210004), China Postdoctoral Science Foundation (Grant No. 2021M700259) and China National Postdoctoral Program for Innovative Talents (Grant No. BX20200017). We thank CompoundTek for SiPh PIC fabrication, and Shenzhen PhotonX Technology Co., Ltd. for laser packaging support.

**Author contributions**
The experiments were conceived by Y.T., with the assistance from F.Y., Z.T., H.S. and M.J. The SiPh PIC was designed by Y.T. The packaging of this on-chip MWP IFM system was conducted by F.Y., Y.T., Y.Z., and Z.G. The results were analyzed by Y.T., F.Y., L.C., and Z.T. All authors participated in writing the manuscript. The project was coordinated by Y.T. and F.Y., under the supervision of X.W.

**Additional information**
Supplementary information is available in the online version of the paper. Reprints and permissions information is available online. Correspondence and requests for materials should be addressed to X.W.

**Conflict of interest**
The authors declare no competing financial interests.